\documentclass{article}

\usepackage{arxiv}

\usepackage{amssymb}
\usepackage{amsmath}
\usepackage{algorithm}
\usepackage{algpseudocode}

\usepackage[hyphens]{url}
\usepackage{float}
\usepackage{mathtools}

\usepackage{booktabs}

\title{Foundation Model for Composite Microstructures: Reconstruction, Stiffness, and Nonlinear Behavior Prediction}

\author{
 Ting-Ju Wei \\
  Department of Civil Engineering\\
  National Taiwan University\\
  Taipei, Taiwan \\
   \And
 Chuin-Shan Chen\thanks{Corresponding author. Email: \texttt{dchen@ntu.edu.tw}} \\
  Department of Civil Engineering\\
  Department of Materials Science and Engineering\\
  National Taiwan University\\
  Taipei, Taiwan \\
}

\begin{document}
\maketitle
\begin{abstract}
We present the Material Masked Autoencoder (MMAE), a self-supervised Vision Transformer pretrained on a large corpus of short-fiber composite images via masked image reconstruction. The pretrained MMAE learns latent representations that capture essential microstructural features and are broadly transferable across tasks. We demonstrate two key applications: (i) predicting homogenized stiffness components through fine-tuning on limited data, and (ii) inferring physically interpretable parameters by coupling MMAE with an interaction-based material network (IMN), thereby enabling extrapolation of nonlinear stress–strain responses. These results highlight the promise of microstructure foundation models and lay the groundwork for future extensions to more complex systems, such as 3D composites and experimental datasets.
\end{abstract}

\keywords{Foundation model \and Composite material \and Transfer learning \and Self-supervised learning \and Masked autoencoder}

\section{Introduction}\label{sec1}

The mechanical performance of composite materials is intrinsically governed by their underlying microstructural architecture. Accurately predicting material properties based on microstructural information is therefore critical for designing high-performance composites. In recent years, machine learning (ML) has emerged as a powerful tool for modeling microstructure–property relationships, enabling rapid property prediction and accelerating the materials design process. For instance, convolutional neural networks (CNNs) have been employed to estimate the effective elastic properties of two-dimensional (2D) checkerboard composites \cite{abueidda2019prediction}. U-Net architectures have been used to predict stress fields in polycrystalline materials, demonstrating the ability to capture spatially complex mechanical responses \cite{mianroodi2021teaching}. Graph neural networks (GNNs) have been integrated into deep material networks for modeling nonlinear composite behavior \cite{jean2024graph}. More recently, transformer-based models have been explored, where pre-trained Vision Transformers were used to capture elastoplastic responses under varied loading conditions \cite{zhongbo2024pre}.

However, despite these successes, most existing ML approaches in materials science are constrained by their reliance on supervised learning with large labeled datasets. Acquiring extensive, high-quality labeled data—whether from experiments or high-fidelity simulations—is expensive and time-consuming. Furthermore, models trained on one type of microstructure or property often fail to generalize to different material systems, especially when faced with unseen microstructural patterns or out-of-distribution cases, which limits their broader applicability~\cite{npjOmee}. In contrast, self-supervised learning techniques have revolutionized fields like natural language processing (NLP) and computer vision by greatly reducing the need for labeled data. Transformer-based language models such as BERT (Bidirectional Encoder Representations from Transformers) and GPT (Generative Pre-trained Transformer) leverage self-supervised objectives (e.g., masked word prediction or next-word prediction) to learn rich representations from massive unlabeled text corpora~\cite{devlin2018bert,  radford2019language}. 

Building upon the success of self-supervised learning, the concept of foundation models has emerged. A foundation model is typically pre-trained on a vast and diverse dataset to learn general-purpose representations that can be adapted to a wide range of downstream tasks~\cite{bommasani2021opportunities, zhou2023comprehensive}. In essence, such models serve as broad, domain-general feature extractors that can be fine-tuned for different problems, marking a shift toward more versatile learning architectures that consolidate knowledge across domains. 

This paradigm has also been applied to computer vision, for example, through masked autoencoders (MAEs) for image data. In an MAE, a transformer-based encoder–decoder model is trained to reconstruct images from which a significant portion of the content has been masked out. Notably, the masked autoencoder introduced by He et al.~\cite{dosovitskiy2020image} employs a Vision Transformer backbone with a lightweight decoder and has achieved performance on par with or even exceeding that of contrastive self-supervised learning methods for image recognition. Such results highlight the potential of MAEs as foundation models in the visual domain.

Despite recent advances in other domains, the application of foundation models in materials science remains at an early stage. We posit that bringing self-supervised foundation model techniques into the materials domain can help address the data scarcity and generalization challenges outlined above. Motivated by this opportunity, we introduce a foundation model specifically designed for composite microstructures, termed the Material Masked Autoencoder (MMAE).

The MMAE is pretrained in a self-supervised fashion on a large corpus of synthetic microstructure images by reconstructing partially masked inputs. This masked reconstruction task enables the model to learn latent representations that encode critical geometric and morphological features of the microstructures. To evaluate the utility and generalizability of these learned representations, we consider two downstream tasks.

First, we predict the homogenized stiffness components of composite representative volume elements (RVEs) by applying transfer learning to the pretrained MMAE using three strategies: linear probing, partial fine-tuning, and end-to-end fine-tuning. These approaches yield accurate predictions even in data-scarce scenarios.

Second, we integrate the pretrained MMAE with a physics-informed surrogate model, the Interaction-based Material Network (IMN), which decouples microstructural geometry from the constitutive behavior of individual phases. Originally developed for composites~\cite{noels2022interaction} and later extended to porous media~\cite{noels2022micromechanics} and polycrystalline materials~\cite{Wei01}, the IMN is grounded in homogenization theory, allowing it to incorporate arbitrary nonlinear constitutive models during online prediction. Its computational efficiency and predictive accuracy have been further validated in recent studies~\cite{wan2024decoding}. The MMAE is end-to-end fine-tuned using only linear elastic data, and its latent features are used to predict IMN parameters. We retain the IMN architecture throughout transfer learning to preserve its theoretical decoupling of geometry and material response. As a result, the trained framework is capable of extrapolating from linear elastic training data to accurately predict nonlinear stress–strain behavior under complex loading conditions.

Through these case studies, we demonstrate that the MMAE learns rich and transferable representations of composite microstructures, enabling both linear property prediction and nonlinear behavior modeling. To our knowledge, this work represents the first application of a foundation model in composite material modeling, where pretrained encoders are leveraged for two distinct downstream tasks: direct stiffness prediction and the parameterization of a physics-informed surrogate model. This approach underscores the versatility of pretrained representations in materials modeling, enabling diverse predictive tasks with improved data efficiency and robust generalization.

The remainder of the paper is organized as follows. Section~\ref{sec2} describes the methodology, including the data generation process and the MMAE model architecture. Section~\ref{sec3} presents the results and discussion. Section~\ref{sec4} offers concluding remarks and outlines potential directions for future work.

\section{Methods}\label{sec2}

\subsection{Self-supervised pre-training using masked autoencoders}

\begin{figure}[h]
    \centering
    \includegraphics[width=\linewidth]{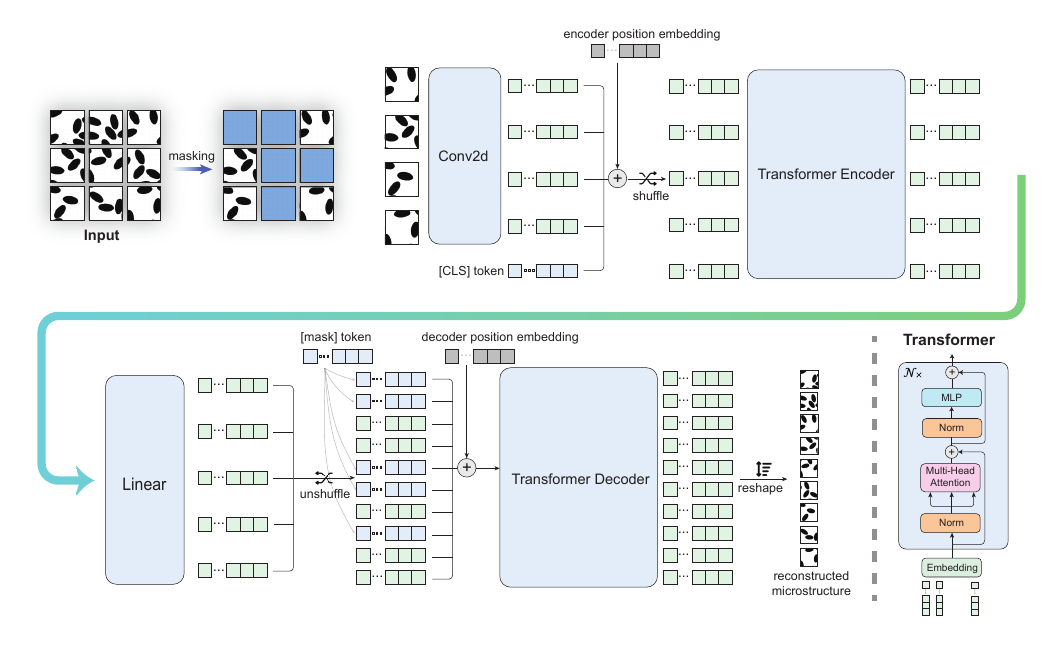}
    \caption{Self-supervised pre-training architecture of the MMAE. The encoder extracts latent features from visible microstructure patches, while the decoder reconstructs the original microstructure image using these extracted embeddings and positional encodings. Light blue regions indicate trainable components; grey blocks are frozen modules; green arrows represent data flow.}
    \label{fig:mmae_architecture}
\end{figure}

The MMAE framework proposed in this work adopted a self-supervised architecture inspired by the MAE introduced by He et al.~\cite{he2022masked}. As illustrated in Fig.~\ref{fig:mmae_architecture}, the MMAE consists of an encoder-decoder structure. The encoder is based on a Vision Transformer (ViT), parameterized by the patch size $\mathcal{P}$, embedding dimension $\mathcal{D}_e$, number of transformer blocks $\mathcal{N}_e$, and number of attention heads per block $\mathcal{H}_e$~\cite{dosovitskiy2020image}. A linear projection layer is used to align the encoder's output dimension with the decoder's input. The decoder shares a similar architecture but with its own parameters: embedding dimension $\mathcal{D}_d$, number of transformer blocks $\mathcal{N}_d$, and number of attention heads $\mathcal{H}_d$. The specific configuration used in this study was: $\mathcal{P} = 16$, $\mathcal{D}_e = 256$, $\mathcal{N}_e = 12$, $\mathcal{H}_e = 4$; 
$\mathcal{D}_d = 128$, $\mathcal{N}_d = 8$, $\mathcal{H}_d = 16$.

Each microstructure was represented as an input image $\mathbf{I} \in \mathbb{R}^{H \times W \times C}$. We used grayscale images ($C=1$) with spatial resolution $H = W = 224$, where each pixel was a binary indicator (0 or 1) representing the phase (e.g., matrix or inclusion). Each image was partitioned into $N_p = \left( \frac{H}{\mathcal{P}} \right)^2 = 196$ non-overlapping patches, where each patch had dimensions of $\mathcal{P} \times \mathcal{P} = 16 \times 16$. These patches were then individually embedded via a convolutional layer and processed by the transformer encoder.

To inject spatial information into the transformer encoder, we employed 2D sinusoidal positional encodings ($\text{PE}$) for both the encoder and decoder inputs. As the self-attention mechanism in transformers is inherently permutation-invariant, such positional encodings are essential to preserve the relative order and layout of image patches. We adopted the fixed 2D extension of the original 1D sinusoidal encoding~\cite{vaswani2017attention}, as commonly used in MAE-based architectures~\cite{dosovitskiy2020image}. Specifically, for a patch grid of size $\mathcal{G} \times \mathcal{G}$ (with $\mathcal{G} = \sqrt{N_p} = 14$ in our case), each spatial coordinate $(x, y)$ was independently encoded using sine and cosine functions at multiple frequencies. For $i = 0, 1, \ldots, \mathcal{D}_e/2 - 1$, the encoding is defined as:
\begin{equation}
\begin{aligned}
\text{PE}_x^{(2i)}(x) &= \sin\left( \frac{x}{10000^{2i / \mathcal{D}_e}} \right), \quad
\text{PE}_x^{(2i+1)}(x) = \cos\left( \frac{x}{10000^{2i / \mathcal{D}_e}} \right), \\
\text{PE}_y^{(2i)}(y) &= \sin\left( \frac{y}{10000^{2i / \mathcal{D}_e}} \right), \quad
\text{PE}_y^{(2i+1)}(y) = \cos\left( \frac{y}{10000^{2i / \mathcal{D}_e}} \right),
\end{aligned}
\end{equation}
where $\text{PE}_x(x), \text{PE}_y(y) \in \mathbb{R}^{\mathcal{D}_e/2}$ represent the positional embeddings for the horizontal and vertical coordinates, respectively. 

In the encoder, we additionally prepended a learnable [CLS] token to the input sequence, following the standard ViT. This [CLS] token is a trainable vector that is initialized randomly and optimized during pre-training. Its role is to serve as a global representation of the input microstructure, capturing holistic spatial characteristics through self-attention.

Since the transformer encoder permits full pairwise attention among all tokens, the [CLS] token attends to every visible patch token, and each patch token similarly attends to the [CLS] token. This bidirectional interaction allows the [CLS] token to aggregate contextual information from the entire set of visible patches, effectively forming a compact latent descriptor of the global microstructural geometry. Consequently, this descriptor can be directly utilized for various downstream prediction tasks.

The positional embedding assigned to the [CLS] token is a zero vector:

\begin{equation}
\text{PE}_{\text{[CLS]}} = \mathbf{0} \in \mathbb{R}^{\mathcal{D}e}.
\end{equation}

The complete positional embedding matrix for the encoder was constructed by stacking all patch embeddings and the [CLS] embedding:

\begin{equation}
\mathbf{PE}_{\text{encoder}} = \text{stack} \left( \text{PE}_{\text{[CLS]}},\, \left\{ \text{concat}\left( \text{PE}_x(x),\, \text{PE}_y(y) \right) \mid (x, y) \in \mathcal{G} \times \mathcal{G} \right\} \right)
\in \mathbb{R}^{(1 + N_p) \times \mathcal{D}_e}
\end{equation}

The transformer decoder, responsible for reconstructing the full image, including masked patches, used the same 2D sinusoidal positional encodings for all patch tokens. While the encoding formulation remained identical, the embeddings were projected into the decoder's latent space with dimension $\mathcal{D}_d$. These positional embeddings provide essential spatial context, enabling the decoder to accurately localize and reconstruct the masked regions.

To implement self-supervised learning, we adopted a random masking strategy. Given a masking ratio $r \in [0,1]$, $r \cdot N_p$ patches were randomly selected and masked, while the remaining served as visible inputs. The encoder processes only the visible patches and produces latent representations that are expected to capture intrinsic morphological characteristics of the input microstructure image, such as fiber volume fraction, aspect ratio, and spatial distribution. These embeddings were then passed to the decoder to reconstruct the pixel values of the masked regions.

The MMAE was trained by minimizing the mean squared error (MSE) between reconstructed and ground-truth pixel values in the masked regions. Specifically, let $\mathbf{I}_{\text{masked}}^{(i)}$ and $\hat{\mathbf{I}}_{\text{masked}}^{(i)}$ represent ground-truth and reconstructed pixels of the $i$-th masked patch, respectively, each of length $d = \mathcal{P} \times \mathcal{P} \times C$. The loss function is defined as

\begin{equation}
    \mathcal{L}_{\text{MMAE}} = \frac{1}{|\mathcal{M}|} \sum_{i \in \mathcal{M}} \frac{1}{d} \left\| \hat{\mathbf{I}}_{\text{masked}}^{(i)} - \mathbf{I}_{\text{masked}}^{(i)} \right\|_2^2,
\end{equation}
where $\mathcal{M}$ denotes the set of masked patch indices and $\lVert \cdot \rVert_2$ is the Euclidean norm.

By leveraging unlabeled microstructure data, this reconstruction-based objective encourages the encoder to learn intrinsic spatial patterns and morphological features, including phase volume fraction, inclusion shape, aspect ratio, and spatial distribution. These representations serve as a transferable basis for downstream tasks such as stiffness prediction and surrogate modeling.

\subsection{Transfer learning for homogenized stiffness prediction}

We evaluated the effectiveness of latent embeddings obtained from the MMAE encoder for predicting homogenized stiffness components. The encoder’s [CLS] token embedding, representing global microstructural information, was passed into a linear prediction head (fully connected layer) to estimate homogenized stiffness values.

\begin{figure}[h]
    \centering
    \includegraphics[width=\linewidth]{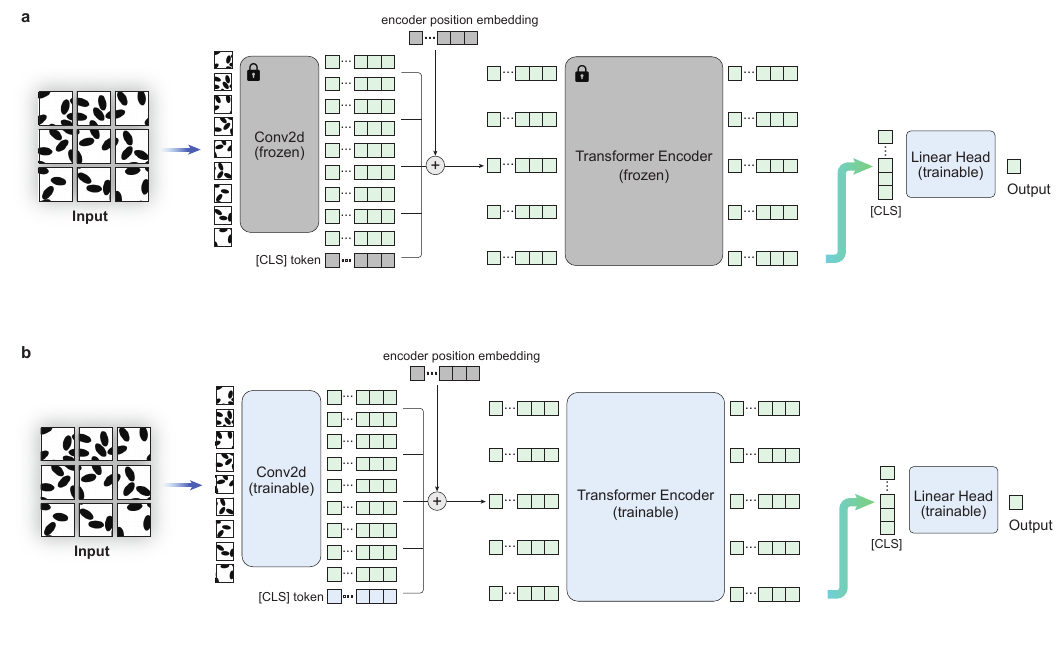}
    \caption{Transfer learning strategies utilizing MMAE embeddings: (a) Linear probing with frozen encoder; (b) Fine-tuning with trainable encoder and prediction head. Grey blocks indicate frozen parameters; light blue blocks indicate trainable parameters; green arrows depict data flow.}
    \label{fig:transfer_learning_arch}
\end{figure}

Two transfer learning approaches, illustrated in Fig.~\ref{fig:transfer_learning_arch}, were investigated:
\begin{itemize}
    \item linear probing: The encoder weights are frozen, and only the linear regression head parameters are trained. This strategy assesses the intrinsic quality of pre-trained features without adapting the encoder.

    \item Fine-tuning: Both encoder and linear head parameters are jointly trained, allowing the encoder's representations to adapt specifically to the stiffness prediction task.

\end{itemize}

\subsubsection{Linear probing}

In linear probing, the MMAE encoder remains entirely frozen, and only the newly introduced linear regression layer (linear head) is trained. This linear head directly uses the encoder-generated [CLS] token embedding to predict homogenized stiffness components ($\bar{C}_{1111}$, $\bar{C}_{2222}$, $\bar{C}_{1212}$). The objective function used is the MSE loss. Linear probing thus isolates and tests the intrinsic predictive power of the MMAE embeddings.

\subsubsection{Fine-tuning}

In fine-tuning, parameters of both the MMAE encoder and linear head are optimized together. Two fine-tuning approaches were explored:
\begin{itemize}
    \item End-to-end fine-tuning: All MMAE encoder parameters are trainable, allowing complete adaptation of the pre-trained model.

    \item Partial fine-tuning: Only the last few transformer blocks of the encoder are updated, while earlier layers remain frozen, providing a more computationally efficient approach that retains general features learned during pre-training.
\end{itemize}

\subsection{Transfer learning for nonlinear behavior prediction}

In addition to predicting linear elastic stiffness, we extend the use of the pretrained MMAE to model the nonlinear mechanical response of composite materials. This is achieved by coupling the MMAE with the IMN, a physics-informed model derived from homogenization theory~\cite{noels2022micromechanics}. The IMN satisfies essential physical constraints, including the Hill–Mandel condition and classical averaging theorems, allowing it to be trained exclusively on linear elastic data while remaining applicable to nonlinear regimes during the online prediction stage~\cite{noels2022micromechanics}.

Conventional IMNs are tailored to fixed RVE geometries and require retraining when the microstructure changes. Recent developments in the Deep Material Networks (DMNs), which share a similar architecture to IMN, have addressed this limitation by incorporating handcrafted microstructure descriptors. These include second-order fiber orientation tensors, which enable generalization across different fiber shapes and distributions~\cite{liu2020intelligent, huang2022microstructure}. Wei et al. further demonstrated the scalability of this approach in large-scale SFRC simulations using LS-DYNA~\cite{wei2023ls}.

In this work, we propose a complementary paradigm based on foundation models, where microstructures are directly represented as 2D images rather than engineered descriptors. This image-based formulation enables self-supervised pretraining on large unlabeled datasets, allowing the model to learn generalizable latent representations of microstructural geometry. The pretrained MMAE encoder is then used to infer IMN parameters. Specifically, the MMAE encoder is fine-tuned end-to-end along with a lightweight linear projection head that maps latent embeddings from microstructure images to a set of IMN parameters. These parameters encode the geometric structure of the input RVE and are used to parameterize the IMN for nonlinear mechanical prediction during the online prediction stage.

This MMAE-IMN framework leverages the complementary strengths of both components: the MMAE provides microstructure-aware representations that generalize across diverse geometries, while the IMN enforces physical consistency and enables accurate prediction of nonlinear mechanical responses.

\subsubsection{Overview of IMN}
The IMN approximates the heterogeneous response of a composite RVE by decomposing it into a limited number of predefined material nodes, each corresponding to distinct subdomains, such as matrix or inclusions. Each material node is associated with a parameter $w_i$ representing the subdomain’s relative contribution (e.g., volume fraction) to the overall mechanical response.

These predefined material nodes are interconnected through interaction nodes, which enforce mechanical compatibility and equilibrium between pairs of subregions. Each interaction node functions as a mechanical interaction mechanism that ensures stress and strain fields satisfy the Hill–Mandel condition for energy consistency. Notably, each interaction node is characterized by an interface direction parameter $\theta_j$, indicating the orientation along which mechanical equilibrium is enforced between its predefined child subdomains. 

By hierarchically aggregating subdomains through these interaction nodes in a binary-tree structure, the IMN constructs a complete material network capable of accurately predicting the homogenized mechanical behavior of composites.

Consequently, for an IMN of depth $N$, the set of all trainable parameters in the 2D formulation is given by:

\begin{equation}\label{eq:imn_parameters}
\mathcal{F}_{\text{IMN}} = \left\{ w_i \mid i = 1, \dots, 2^N  \right\} \cup \left\{ \theta_j \mid j = 1, \dots, 2^N - 1 \right\}, \quad \text{with} \quad |\mathcal{F}_{\text{IMN}}| = 2^{N+1} - 1.
\end{equation}

Here, $w_i$ represents the volume fraction associated with each leaf material node, and $\theta_j$ encodes the in-plane orientation of the interface at each interaction node. This parameterization is specific to 2D RVEs under plane strain assumptions. To extend the IMN to 3D microstructures, this formulation must be generalized to accommodate additional geometric degrees of freedom. In particular, each interaction interface would require two angular parameters (e.g., $\theta$ and $\varphi$) to define the orientation of its normal vector in three-dimensional space.

Given a pair of constituent stiffness tensors $(\mathbf{C}^{p1}, \mathbf{C}^{p2})$, the IMN performs a closed-form homogenization to compute the effective stiffness tensor $\bar{\mathbf{C}}$ as:

\begin{equation}\label{eq:IMN_mapping}
(\mathbf{C}^{p1}, \mathbf{C}^{p2}) \xmapsto[\mathcal{F}_{\text{IMN}}]{} \bar{\mathbf{C}}.
\end{equation}
The detailed formulation of this homogenization process can be found in~\cite{noels2022interaction,noels2022micromechanics}.

During the offline training phase, the goal is to learn the optimal IMN parameters $\mathcal{F}_{\text{IMN}}$ such that the homogenized stiffness prediction $\bar{\mathbf{C}}$ closely matches the ground-truth stiffness tensor $\bar{\mathbf{C}}^{\text{DNS}}$ computed from direct numerical simulations (DNS). Given a dataset of constituent stiffness pairs $(\mathbf{C}^{p1}, \mathbf{C}^{p2})$ and corresponding target homogenized stiffness tensors $\bar{\mathbf{C}}^{\text{DNS}}$, the IMN is trained via gradient-based optimization to minimize the following relative Frobenius norm loss:

\begin{equation}
    \mathcal{L}_{\text{IMN}} = \frac{1}{N_{\text{dataset}}} \sum_{s=1}^{N_{\text{dataset}}} \frac{\lVert \bar{\mathbf{C}}^{\text{DNS}}_s - \bar{\mathbf{C}}_s(\mathbf{C}^{p1}_s, \mathbf{C}^{p2}_s |\mathcal{F}_{\text{IMN}} ) \rVert_{\text{F}}}{\lVert \bar{\mathbf{C}}^{\text{DNS}}_s \rVert_{\text{F}}},
\label{eq:imn_loss}
\end{equation}
where $N_{\text{dataset}}$ is the total number of training samples, and $\lVert \cdot \rVert_{\text{F}}$ denotes the Frobenius norm.

\subsubsection{Offline end-to-end fine-tuning}

Traditionally, IMN parameters $\mathcal{F}_{\text{IMN}}$ are individually optimized for each microstructure through offline training. Here, we propose a novel approach where the IMN parameters are directly inferred from the microstructure image using MMAE embeddings, significantly improving efficiency and generalizability.

\begin{figure}[h]
    \centering
    \includegraphics[width=1\linewidth]{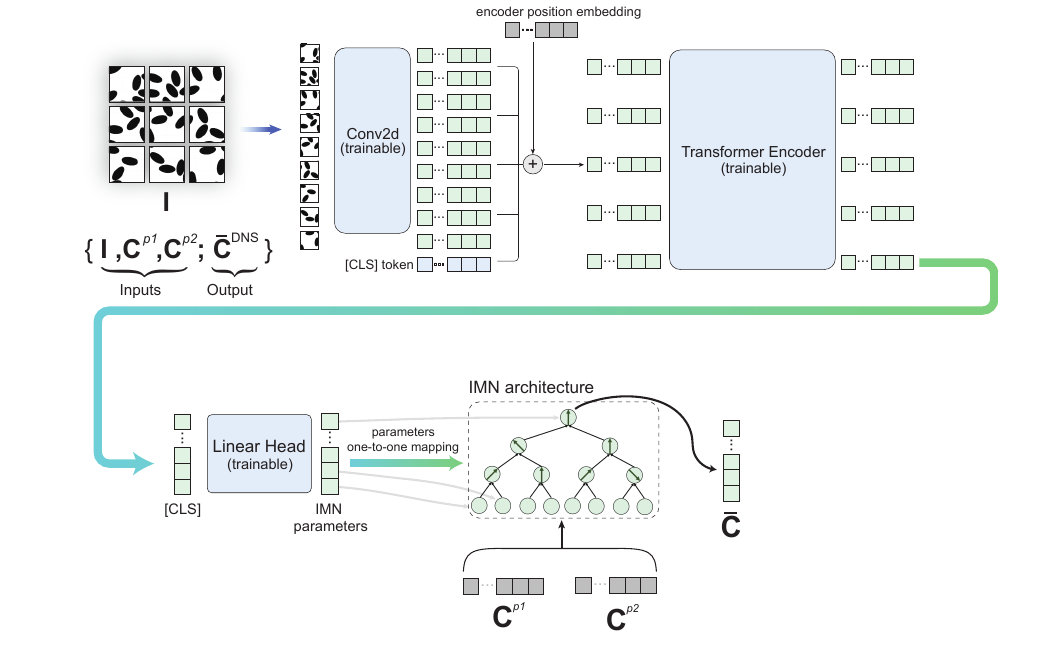}
    \caption{End-to-end transfer learning framework from MMAE to IMN. Each input microstructure image $\mathbf{I}$ is encoded by the pre-trained MMAE, and its [CLS] token serves as a latent representation. A linear projection head maps this embedding to predict the IMN parameter set $\mathcal{F}_{\text{IMN}}(\mathbf{I})$. The IMN then uses these parameters, along with the stiffness matrices of phase 1 and phase 2 ($\mathbf{C}^{p1}$, $\mathbf{C}^{p2}$), to compute the predicted homogenized stiffness tensor $\bar{\mathbf{C}}$. Grey blocks indicate frozen (non-trainable) modules, light blue blocks denote trainable components, and green arrows represent the data flow throughout the pipeline.}
    \label{fig:fm_imn_framework}
\end{figure}

As illustrated in Fig.~\ref{fig:fm_imn_framework}, the proposed framework integrates the pretrained MMAE encoder with the IMN. The process begins by encoding an input microstructure image $\mathbf{I}$ with MMAE, from which the latent representation is extracted via the [CLS] token. A lightweight linear projection head then maps this embedding to a predicted parameter set $\mathcal{F}_{\text{IMN}}(\mathbf{I})$.

Crucially, we preserve the original IMN architecture and enforce a strict one-to-one correspondence between the projection outputs and the IMN parameters defined in Eq.~\eqref{eq:imn_parameters}. This ensures that the inferred parameters remain physically interpretable and fully compatible with the underlying homogenization theory, including the Hill–Mandel condition and averaging theorems.

The predicted parameter set $\mathcal{F}_{\text{IMN}}(\mathbf{I})$, together with the stiffness tensors of the constituent phases $(\mathbf{C}^{p1}, \mathbf{C}^{p2})$, is then passed to the IMN to compute the homogenized stiffness tensor $\bar{\mathbf{C}}$.

This mapping can be succinctly expressed as:
\begin{equation}\label{eq:end_to_end_mapping}
(\mathbf{I}, \mathbf{C}^{p1}, \mathbf{C}^{p2}) \xmapsto[\mathcal{F}_{\text{IMN}}(\mathbf{I})]{} \bar{\mathbf{C}}.
\end{equation}

The entire MMAE-IMN framework is trained end-to-end by minimizing the discrepancy between the IMN-predicted stiffness $\bar{\mathbf{C}}$ and ground truth stiffness $\bar{\mathbf{C}}^{\text{DNS}}$:

\begin{equation}
    \mathcal{L}_{\text{MMAE-IMN}} = \frac{1}{N_{\text{dataset}}} \sum_{s=1}^{N_{\text{dataset}}} \frac{\lVert \bar{\mathbf{C}}^{\text{DNS}}_s - \bar{\mathbf{C}}_s(\mathbf{C}^{p1}_s, \mathbf{C}^{p2}_s | \mathcal{F}_{\text{IMN}}(\mathbf{I}_s) ) \rVert_{\text{F}}}{\lVert \bar{\mathbf{C}}^{\text{DNS}}_s \rVert_{\text{F}}},
\label{eq:nmse}
\end{equation}

\subsubsection{Online nonlinear prediction using inferred IMN}

\begin{figure}[h]
    \centering
    \includegraphics[width=1\linewidth]{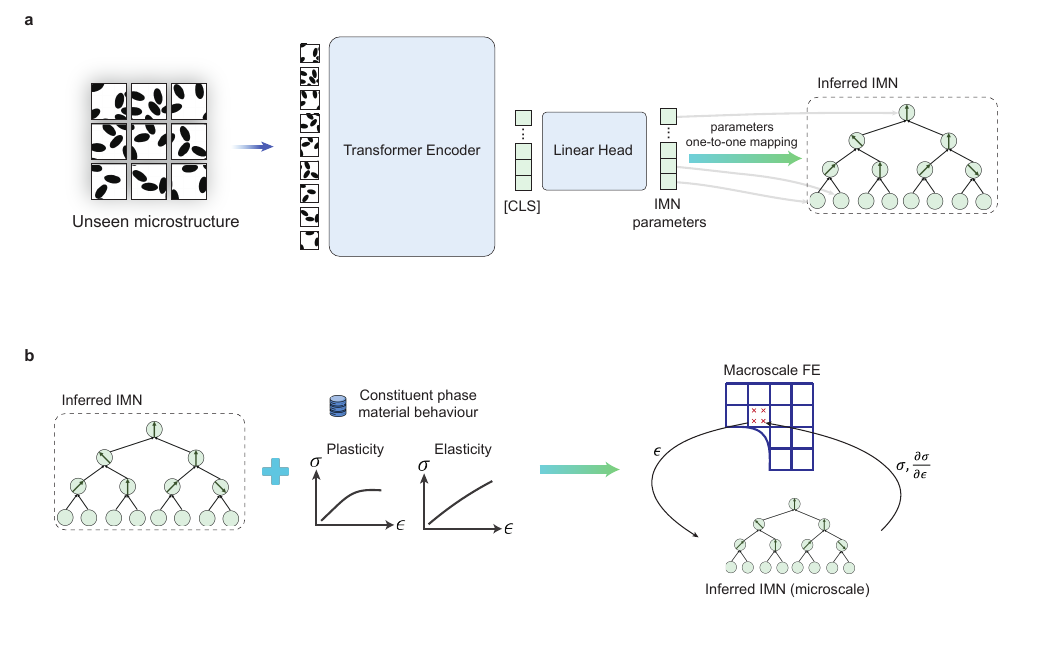}
    \caption{Schematic of online nonlinear prediction using the inferred IMN. (a) A previously unseen microstructure image is encoded by the MMAE and projected to a set of IMN parameters. (b) The inferred IMN, when combined with phase-specific constitutive models (e.g., elasticity or plasticity), enables nonlinear response prediction.}
    \label{fig:fm_imn_online}
\end{figure}

After completing offline training, the MMAE-IMN framework enables efficient online prediction of nonlinear mechanical responses for previously unseen microstructures. Given a new microstructure image $\mathbf{I}$, the MMAE encoder and its linear projection head jointly infer the IMN parameter set $\mathcal{F}_{\text{IMN}}(\mathbf{I})$ through a single forward pass. The inferred IMN, fully parameterized by these predicted values, acts as a surrogate model that captures the microstructure-dependent geometry, as shown in Fig.~\ref{fig:fm_imn_online}(a).

The inferred IMN can be used as a standalone material surrogate model, independent of the MMAE encoder. Nonlinear mechanical behavior is simulated by assigning constitutive models, such as elasticity or plasticity, to the constituent phases at the material node level, as shown in Fig.~\ref{fig:fm_imn_online}(b). By leveraging the MMAE to extract geometry-aware representations from microstructure images and using the IMN to simulate nonlinear mechanical responses at the RVE scale, the framework enables accurate predictions for previously unseen microstructures and loading conditions without requiring additional retraining. The algorithm of the IMN for nonlinear response prediction is detailed in~\cite{noels2022interaction,noels2022micromechanics}.

\subsection{Dataset generation}

\subsubsection{Generation of microstructure}
\label{sec:Generation of microstructure}

Synthetic microstructures were generated for both MMAE pre-training and downstream tasks using the random sequential adsorption (RSA) method~\cite{feder1980random}. RSA is a stochastic algorithm that sequentially places particles into a domain without overlap, enabling the creation of composite microstructures with diverse inclusion counts, shapes, sizes, and orientations. This method captures the inherent randomness of real-world material systems and facilitates the creation of statistically representative RVEs. A detailed pseudocode outlining this generation process is provided in \ref{sec:Microstructure_generation:appendix}.

For short-fiber composites, each RVE contains elliptical inclusions randomly embedded within a matrix. The geometry of each microstructure is defined by three independent descriptors:

\begin{enumerate}
    \item Number of Particles (\( N_p \)): Number of inclusions per RVE, sampled uniformly between 15 and 35.
    \item Aspect Ratio (\( A_r \)): Ratio of the major to minor axes of each elliptical inclusion, ranging from 1 to 4.
    \item Volume Fraction (\( v_f \)): Volume percentage of inclusions, sampled between 10\% and 40\%.
\end{enumerate}

To ensure broad coverage of the design space, we applied Latin Hypercube Sampling (LHS) across the \((N_p, A_r, v_f)\) space~\cite{mckay2000comparison}. For each sampled triplet, the semi-axes \((a, b)\) of the ellipses were analytically computed to satisfy the target volume fraction given the number and aspect ratio of inclusions. All fibers within a single RVE share the same shape and size, while their positions and orientations are assigned independently. Specifically, positions are sampled via uniform random placement and orientation angles are drawn from a uniform distribution over \([0, \pi)\).

The fiber size was determined by $N_p$ and $v_f$, and their orientations were assigned randomly to reflect stochastic microstructural variability.

We constructed three datasets using this procedure:

\begin{enumerate}
    \item Pre-training dataset: 100,000 grayscale images of short-fiber microstructures for self-supervised MMAE pre-training. This dataset spans a wide variety of inclusion geometries and distributions, enabling robust feature learning.

    \item Homogenized stiffness dataset: 5,000 short-fiber microstructures paired with DNS-derived components of the homogenized stiffness matrix (\(\bar{C}_{1111}\), \(\bar{C}_{2222}\), \(\bar{C}_{1212}\)). This dataset is used for supervised training in the downstream task of predicting linear stiffness components.

    \item IMN dataset: 300 short-fiber microstructures, each paired with 300 distinct combinations of phase stiffness matrices \((\mathbf{C}^{p1}_s, \mathbf{C}^{p2}_s)\), resulting in a total of 90,000 DNS simulations. Each simulation yields a homogenized stiffness matrix \(\bar{\mathbf{C}}^{\text{DNS}}_s\). This dataset is used for supervised training in the downstream task, where the model learns to predict the parameters of the IMN. These parameters are subsequently used for nonlinear behavior prediction during online prediction stage.

\end{enumerate}

To assess generalization across different inclusion geometries, we additionally generated a 4th dataset of 5,000 RVEs containing circular inclusions, with volume fractions \(v_f\) uniformly sampled between 10\% and 40\%.

\subsubsection{Dataset for homogenized stiffness prediction}
\label{sec:Dataset for homogenized stiffness prediction}

To support transfer learning from MMAE for predicting individual components of the homogenized stiffness matrix (\(\bar{C}_{1111}\), \(\bar{C}_{2222}\), \(\bar{C}_{1212}\)), we constructed a dataset comprising 5{,}000 distinct short-fiber RVEs with varying inclusion geometries. Each RVE was simulated under a fixed material configuration, where the following linear elastic properties were uniformly assigned to the matrix and inclusion phases:

\begin{itemize}
    \item Matrix: Young's modulus \( E = 100 \), Poisson's ratio \( \nu = 0.30 \).
    \item Inclusions: Young's modulus \( E = 500 \), Poisson's ratio \( \nu = 0.19 \).
\end{itemize}

DNS was performed using the RVE module in LS-DYNA, employing the \texttt{*RVE\_ANALYSIS\_FEM} keyword to impose periodic boundary conditions and compute the effective mechanical response~\cite{lsdyna_manual}. For each RVE, the three stiffness components were extracted from the resulting stress–strain curves and used as targets for supervised regression. The final dataset comprises 5{,}000 labeled instances, each pairing a unique microstructure with its corresponding stiffness values. The data were randomly split into 80\% for training and 20\% for validation.

\subsubsection{Dataset for IMN transfer learning}

To support transfer learning from MMAE to IMN, we constructed a dataset that enables the IMN to learn the mapping from microstructure images and constituent phase stiffness matrices to homogenized stiffness matrices. The dataset consists of 300 short-fiber RVEs, each paired with 300 distinct combinations of phase stiffness matrices $(\mathbf{C}^{p1}_s, \mathbf{C}^{p2}_s)$ for the matrix and inclusion phases. These combinations were generated using the sampling strategy proposed in our previous work, designed to ensure broad coverage of phase contrast variability~\cite{jean2024graph}.

More specifically, both constituent phases are modeled as isotropic linear elastic materials characterized by Young’s modulus $E$ and Poisson’s ratio $\nu$. To balance numerical stability with contrast diversity, we fixed the matrix stiffness to $\ln(E_1) = 0$ and defined the sampling space for the inclusion stiffness as $\ln(E_2) \in [-5,\ 5]$. This corresponds to a stiffness contrast of up to approximately two orders of magnitude ($\sim$ 148×), covering the range commonly encountered in engineering composites. The Poisson’s ratios $\nu_1$ and $\nu_2$ were independently sampled from $[0,\ 0.5]$ to capture a broad spectrum of material compressibility.

Each RVE–material pair is then simulated using the DNS setup described in Section~\ref{sec:Dataset for homogenized stiffness prediction}, which includes periodic displacement boundary conditions. Specifically, we impose small prescribed macroscopic displacement gradients along the $x$, $y$, and shear directions (each with a magnitude of $1.0 \times 10^{-8}$). The resulting stress–strain responses are used to compute the homogenized stiffness matrix \(\bar{\mathbf{C}}^{\text{DNS}}_s\), which serves as the supervised learning target. 

This procedure yields a total of 90{,}000 labeled instances. The dataset is partitioned at the RVE level, with 240 RVEs used for training and the remaining 60 RVEs reserved for validation.

\section{Results and Discussion}\label{sec3}

This section comprehensively evaluates the proposed MMAE framework across tasks of increasing complexity. We assess reconstruction quality under different masking ratios to analyze the learned latent representations. We then assess the transferability of these representations for predicting homogenized stiffness components using linear probing, partial fine-tuning, and end-to-end fine-tuning. We further examine the effects of masking ratios, dataset size, and microstructural morphology. Finally, we demonstrate the model’s ability to predict nonlinear mechanical behavior by fine-tuning it to infer IMN parameters, enabling physics-informed extrapolation under arbitrary loading conditions. Collectively, these results highlight the MMAE’s effectiveness as a generalizable foundation model for microstructure-informed materials modeling.

\subsection{Pre-trained MMAE reconstruction performance}

To qualitatively assess the representational capacity of the pre-trained MMAE model, we examine its reconstruction performance under different pretraining conditions. As the masking ratio during pretraining may influence downstream task performance, we first evaluate how well the MMAE can reconstruct composite microstructures under challenging masking scenarios before proceeding to downstream analyses. Specifically, we focus on high masking ratios of $75\%$ and $85\%$, which impose more demanding reconstruction conditions and better reflect the MMAE’s capacity for feature learning.

\begin{figure}[H]
    \centering
    \includegraphics[width=0.736\linewidth]{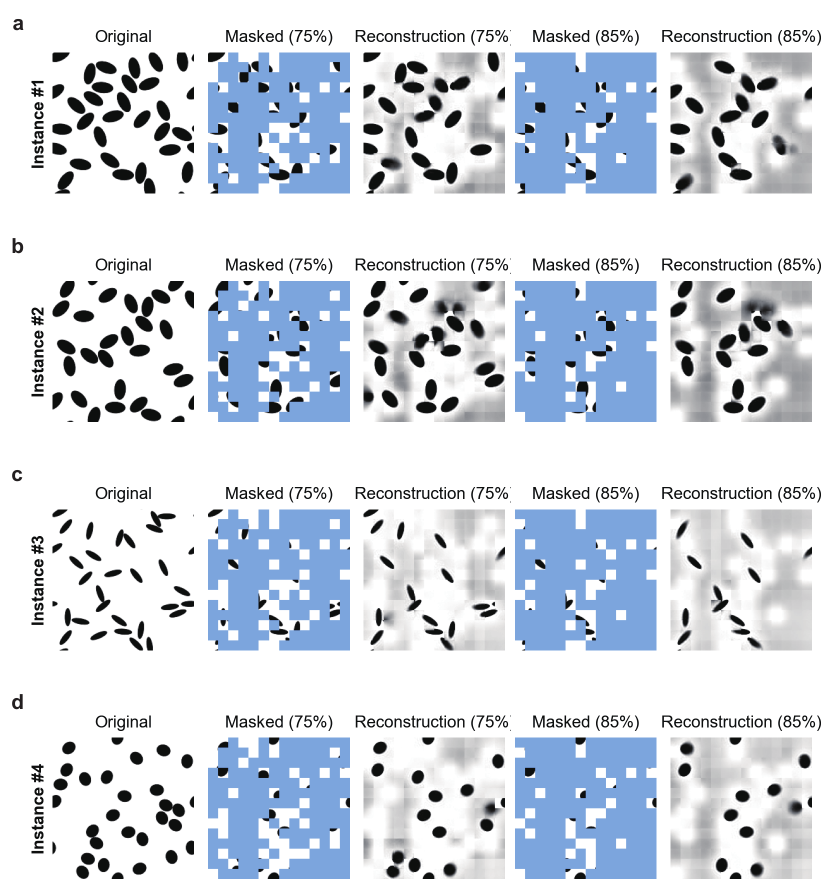}
    \caption{Microstructure reconstruction results using the pretrained MMAE under two masking ratios. Each row ((a)–(d)) corresponds to a randomly selected microstructure from the pretraining dataset, differing in key morphological descriptors such as aspect ratio, volume fraction, and number of inclusions. From left to right: original image, masked input, and reconstruction under 75\% masking; followed by masked input and reconstruction under 85\% masking.}
    \label{fig:reconstruction_shortfiber}
\end{figure}

We randomly select four training instances from the pretraining dataset, each exhibiting distinct morphological characteristics in terms of aspect ratio, volume fraction, and inclusion count, as described in Section~\ref{sec:Generation of microstructure}. These examples are denoted as Instance~\#1 to~\#4 and presented in Fig.~\ref{fig:reconstruction_shortfiber}(a)--(d). Despite the high masking ratios, the reconstructed microstructures visually resemble their original configurations. In this context, we refer to the reconstructions as semantically plausible, meaning that the predicted inclusion shapes and sizes are consistent with those observed in the visible patches, and the overall morphology appears realistic and coherent.

\begin{figure}[H]
    \centering
    \includegraphics[width=0.736\linewidth]{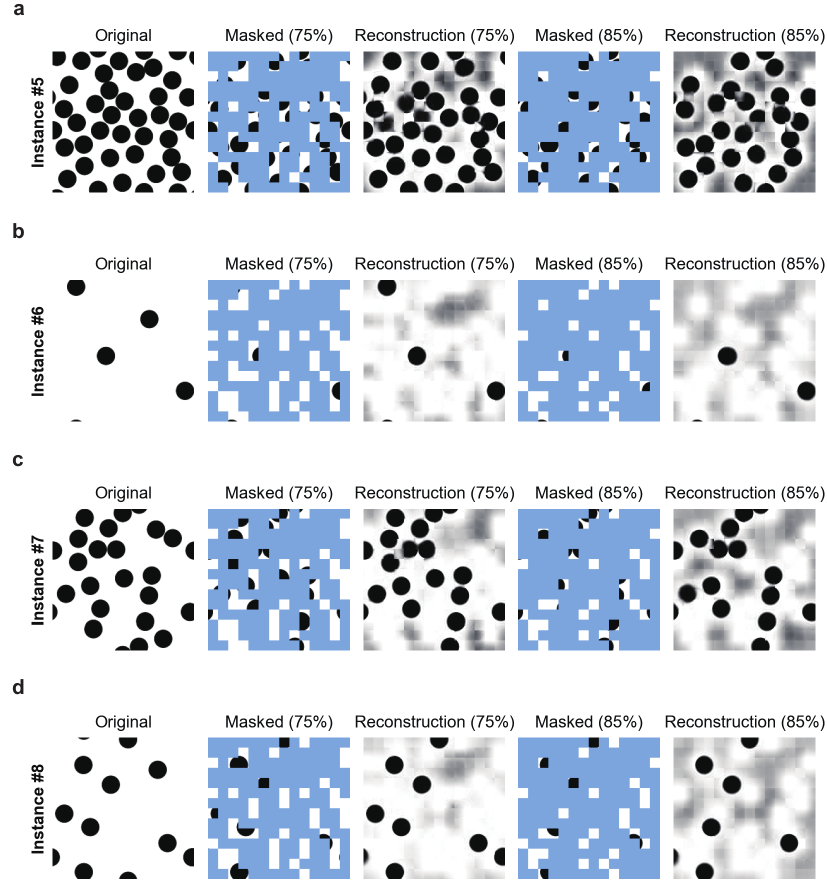}
    \caption{Microstructure reconstruction results using the pretrained MMAE under two masking ratios. Each row ((a)–(d)) corresponds to a circular-inclusion composite with a distinct volume fraction, achieved by varying the number of inclusions while keeping their size constant. From left to right: original image, masked input, and reconstruction under 75\% masking; followed by masked input and reconstruction under 85\% masking.}
    \label{fig:reconstruction_circular}
\end{figure}

Additionally, to further examine the MMAE’s ability to capture geometric variability within the training distribution, we design four synthetic circular-inclusion composites that represent a geometric special case of the short-fiber morphology. Although such circular configurations are not explicitly present in the pretraining dataset, they are geometrically consistent with the short-fiber design space, corresponding to an aspect ratio of one and uniform inclusion size. By varying the number of inclusions, we generate instances with different volume fractions while preserving shape and scale. These examples, labeled as Instance~\#5 to~\#8 in Fig.~\ref{fig:reconstruction_circular}(a)--(d), also yield semantically plausible reconstructions, even under the more challenging $85\%$ masking ratio. This indicates that the MMAE effectively learns to reconstruct structurally reasonable microstructures based on partial observations, even for geometries not explicitly seen during pretraining.

\subsection{Transfer learning for homogenized stiffness components}

We evaluated the effectiveness of the learned latent representations for predicting homogenized stiffness components ($\bar{C}_{1111}$, $\bar{C}_{2222}$, $\bar{C}_{1212}$) using both short-fiber and circular-inclusion composite datasets. We employed three strategies: linear probing, end-to-end fine-tuning, and partial fine-tuning.

\subsubsection{Linear probing}

To evaluate the representational quality of the pretrained encoder, we conducted linear probing using the [CLS] token output from MMAE. A lightweight linear regressor with 257 trainable parameters was trained on top of the frozen embeddings to predict the homogenized stiffness components. Since the regressor lacks the capacity to adapt or refine the latent features, its performance directly reflects the quality and transferability of the representations learned during self-supervised pretraining.

\begin{figure}[h]
    \centering
    \includegraphics[width=\linewidth]{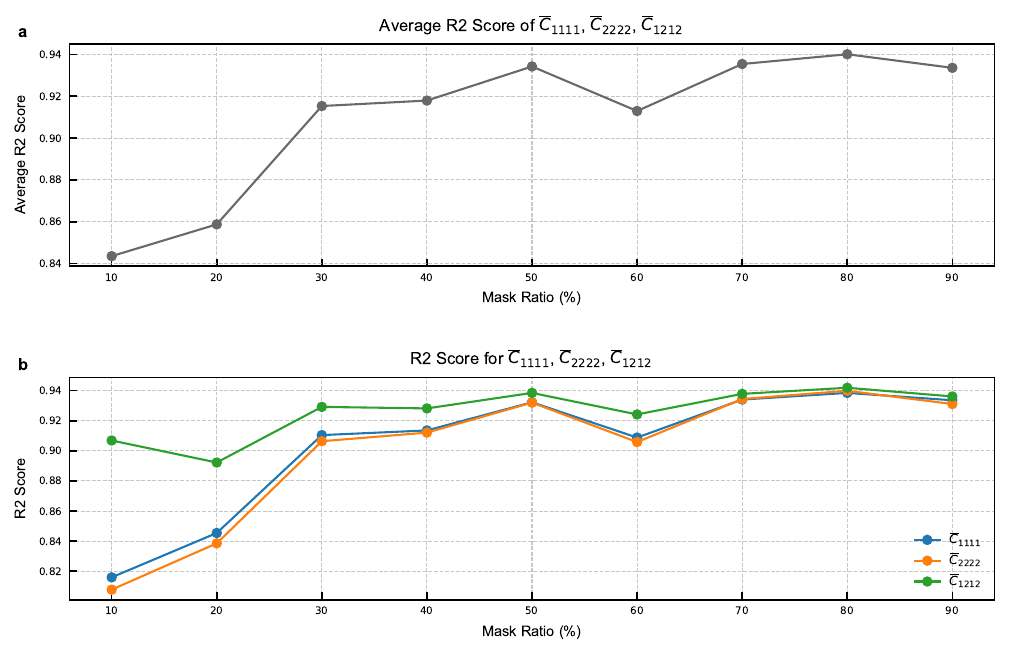}
    \caption{Linear probing performance on the short-fiber composite dataset. (a) Average validation $\mathrm{R}^2$ scores across all stiffness components vs. masking ratios. (b) $\mathrm{R}^2$ scores for each individual stiffness component vs. masking ratios.}
    \label{fig:linear_probing_SFC}
\end{figure}

Fig.~\ref{fig:linear_probing_SFC} presents the results on the short-fiber composite dataset. The validation $\mathrm{R}^2$ scores drop significantly as the masking ratio decreases from $30\%$ to $10\%$, while higher masking ratios yield improved performance, peaking at $\mathrm{R}^2=0.94$ under $80\%$ masking. This trend reflects the impact of masking on representational learning: when too few patches are masked, the reconstruction task becomes overly simple, enabling the model to rely on visible patch interpolation rather than learning abstract and microstructurally informative features. Consequently, the resulting representations exhibit limited generalization, which is particularly evident in linear probing.

\begin{figure}[h]
    \centering
    \includegraphics[width=\linewidth]{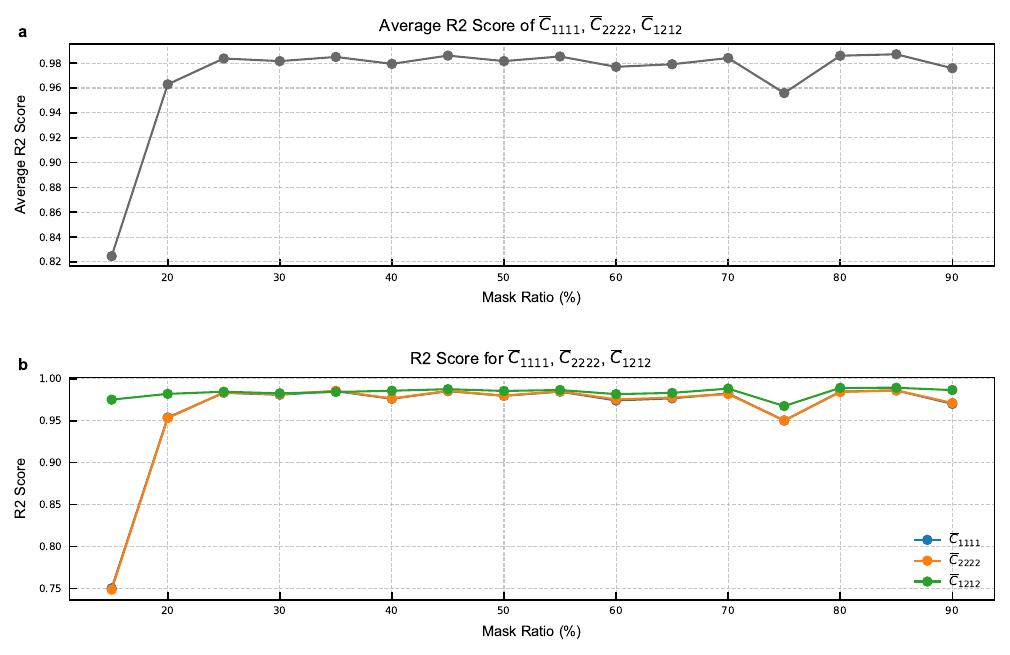}
    \caption{Linear probing performance on the circular-inclusion composite dataset. (a) Average validation $\mathrm{R}^2$ scores vs. masking ratios. (b) $\mathrm{R}^2$ scores for each stiffness component vs. masking ratios.}
    \label{fig:linear_probing_circle}
\end{figure}

Fig.~\ref{fig:linear_probing_circle} shows the results for the circular-inclusion dataset. While this morphology was not explicitly included during pretraining, it can be regarded as a geometric special case within the short-fiber composite family. The model maintains high performance ($\mathrm{R}^2 > 0.975$) across most masking ratios, with only a slight dip at $75\%$. These results indicate that the learned representations remain effective for this specific configuration.

In summary, higher masking ratios during self-supervised pretraining facilitate the learning of more abstract and transferable representations by enforcing stronger representational constraints. This effect is consistent with observations reported in both computer vision and scientific domains~\cite{he2022masked, lee2024max, georgescu2023masked}.

\subsubsection{End-to-end fine-tuning}

To further explore the MMAE’s representational capacity, we performed end-to-end fine-tuning on both datasets. Unlike linear probing, this approach updates all encoder and prediction head parameters, allowing full adaptation to the downstream task.

\begin{figure}[h]
    \centering
    \includegraphics[width=\linewidth]{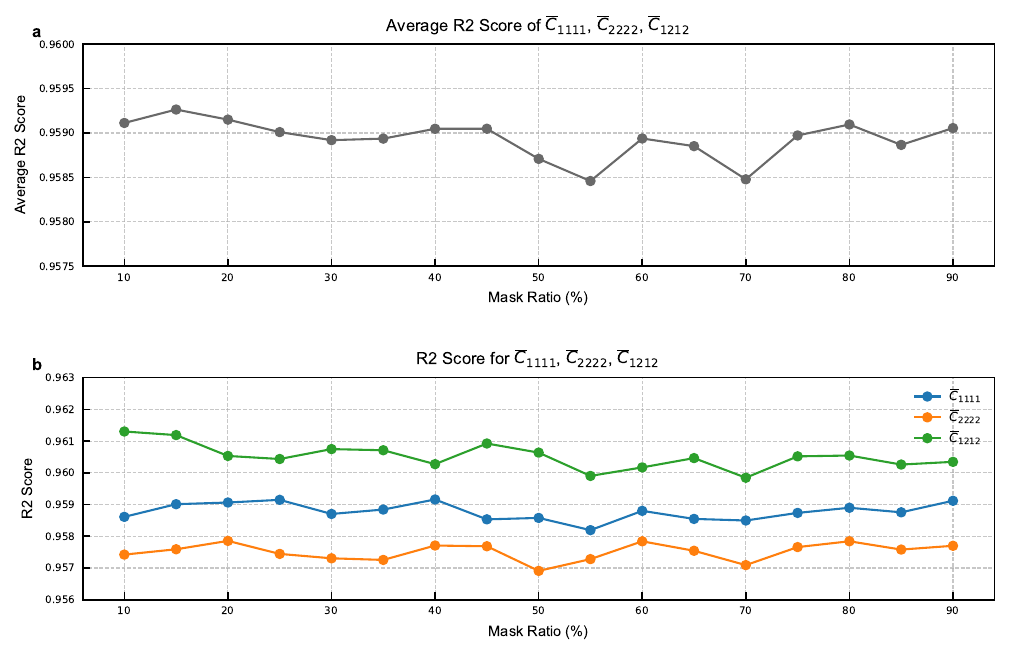}
    \caption{End-to-end fine-tuning performance on the short-fiber composite dataset. (a) Average validation $\mathrm{R}^2$ scores across all stiffness components vs. masking ratios. (b) $\mathrm{R}^2$ scores for each individual stiffness component vs. masking ratios.}
    \label{fig:fine_tuning_SFC}
\end{figure}

As shown in Fig.~\ref{fig:fine_tuning_SFC}, fine-tuning on the short-fiber dataset achieves consistently high $\mathrm{R}^2$ scores, largely insensitive to the pre-training masking ratios. The performance remains largely insensitive to the masking ratio, with the best configuration achieving $\mathrm{R}^2 > 0.959$. This robustness indicates that the model can effectively adapt to the target task by compensating for suboptimal pre-training through joint parameter optimization.

\begin{figure}[h]
    \centering
    \includegraphics[width=\linewidth]{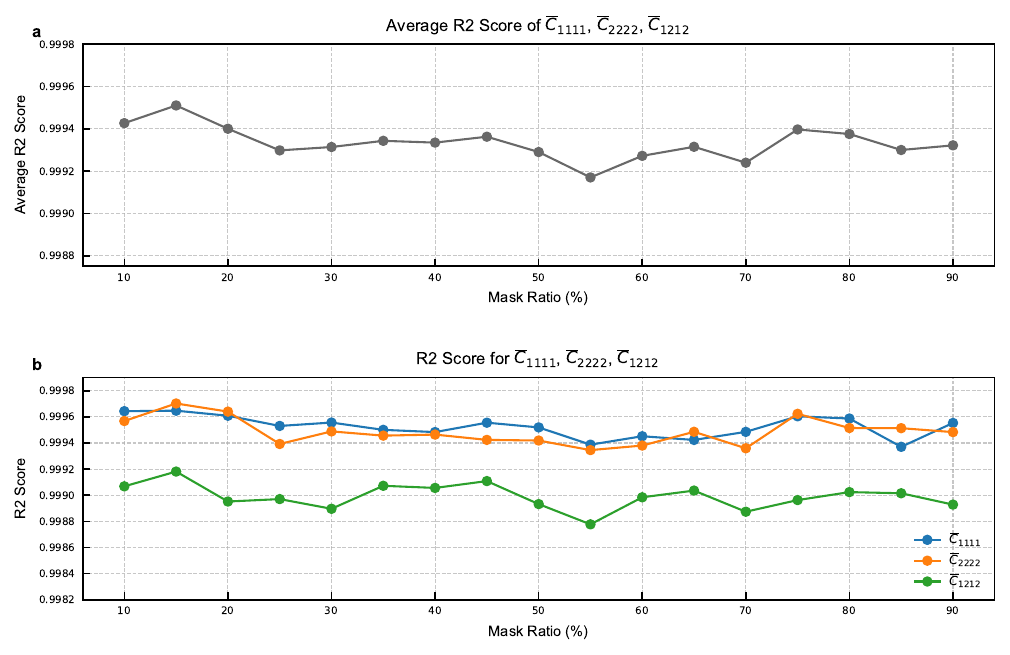}
    \caption{End-to-end fine-tuning performance on the circular-inclusion composite dataset. (a) Average validation $\mathrm{R}^2$ scores vs. masking ratios. (b) $\mathrm{R}^2$ scores for each stiffness component vs. masking ratios.}
    \label{fig:fine_tuning_CIC}
\end{figure}

Similar trends are observed on the circular-inclusion dataset (Fig.~\ref{fig:fine_tuning_CIC}). Despite being trained on different morphologies, the fine-tuned model achieves $\mathrm{R}^2$ scores exceeding 0.999 across all masking ratios, confirming the adaptability of the pre-trained MMAE to diverse geometries.

These results demonstrate that, although the choice of masking ratio plays a critical role in linear probing, its impact is substantially reduced in end-to-end fine-tuning. By jointly updating the encoder and prediction head, the model can refine its representations and compensate for differences in pretraining conditions, thereby achieving consistently high predictive accuracy across masking ratios.

\subsubsection{Partial fine-tuning}

To explore the trade-off between model capacity and training efficiency, we conducted partial fine-tuning experiments. Only the last few transformer blocks of the MMAE encoder were updated, while the early blocks remained frozen.

\begin{figure}[h]
    \centering
    \includegraphics[width=\linewidth]{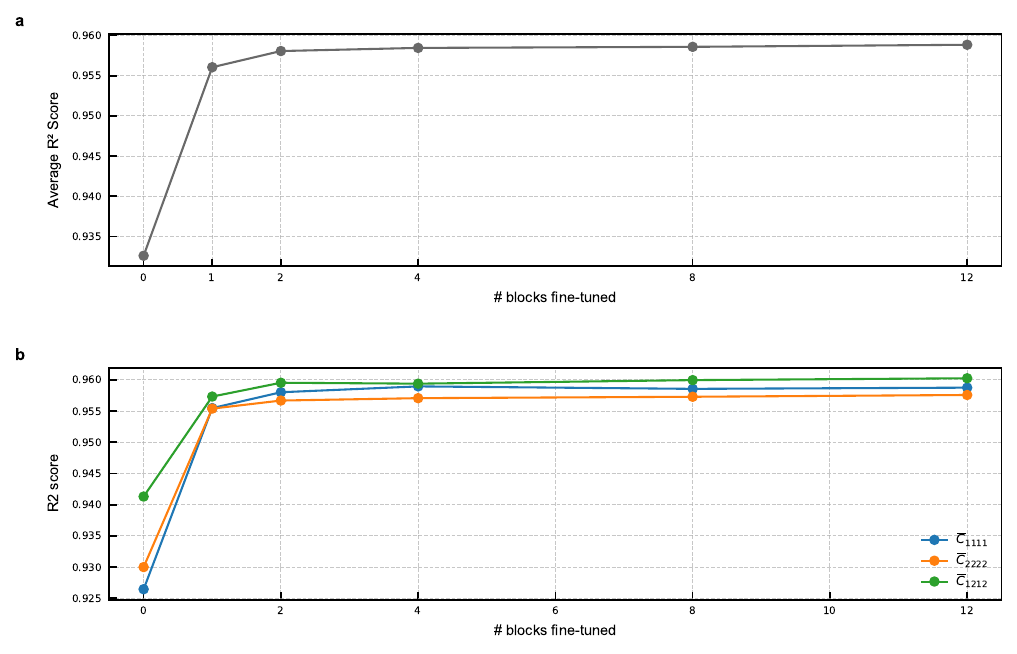}
    \caption{Partial fine-tuning results on the short-fiber dataset (masking ratio = $85\%$). (a) Average $\mathrm{R}^2$ scores vs. number of transformer blocks fine-tuned. (b) $\mathrm{R}^2$ scores for each component vs. the number of transformer blocks fine-tuned.}
    \label{fig:partial_fine_tuning_SFC}
\end{figure}

As shown in Fig.~\ref{fig:partial_fine_tuning_SFC}, performance steadily improves as more blocks are fine-tuned. Updating just the final two blocks yields a substantial performance gain, approaching that of full fine-tuning. This suggests that the early transformer blocks capture general-purpose microstructural features, whereas the later blocks are primarily responsible for task-specific adaptation.

These findings indicate that the mapping from microstructure to homogenized stiffness exhibits relatively low complexity, such that adapting only a small subset of parameters suffices to achieve high accuracy. This property is desirable for foundation models, as it enables efficient transfer with reduced risk of overfitting and lower computational cost.

To quantify the efficiency–performance trade-off across different transfer learning strategies, we compared the training costs associated with linear probing, partial fine-tuning (with varying numbers of transformer blocks), and end-to-end fine-tuning. The results are summarized in Table~\ref{tab:tl_cost_comparison}, which reports the number of trainable parameters and the corresponding training time for each configuration. All experiments were conducted on the short-fiber composite dataset using 5{,}000 training instances to predict the homogenized stiffness component $\bar{C}_{1111}$, with a batch size of 50 and a total of 600 training epochs on an NVIDIA V100 GPU (32~GB).

An important observation from these results relates to computational efficiency. Although the number of trainable parameters increases significantly as more transformer blocks are fine-tuned, the training times remain remarkably similar across different configurations. This consistency arises because the batch size was fixed, and the GPU memory (32 GB) was not fully utilized in any of the tested scenarios. Consequently, the wall-clock training time was primarily limited by data loading and per-epoch overhead rather than GPU-bound computation.

\begin{table}[htbp]
\centering
\caption{Training cost comparison across different transfer learning strategies.}
\label{tab:tl_cost_comparison}
\begin{tabular}{lcc}
\toprule
Configuration & Trainable parameters & Training time \\
\midrule
Linear probing            & 257        & 1h 30m 40s \\
1 block fine-tuned        & 790,017    & 2h 44m 25s \\
2 blocks fine-tuned       & 1,579,777  & 2h 45m 28s \\
4 blocks fine-tuned       & 3,159,297  & 2h 45m 45s \\
8 blocks fine-tuned       & 6,318,337  & 2h 46m 29s \\
End-to-end fine-tuning    & 9,594,369  & 2h 52m 43s \\
\bottomrule
\end{tabular}
\end{table}

\newpage

\subsubsection{Impact of dataset size in transfer learning}

To assess the model’s data efficiency, we evaluated performance as a function of dataset size using the short-fiber composite dataset and an MMAE pre-trained at an $85\%$ masking ratio. In these experiments, we adopted an end-to-end fine-tuning strategy, where both the MMAE encoder and the linear prediction head were jointly optimized using the target data.

\begin{figure}[h]
    \centering
    \includegraphics[width=\linewidth]{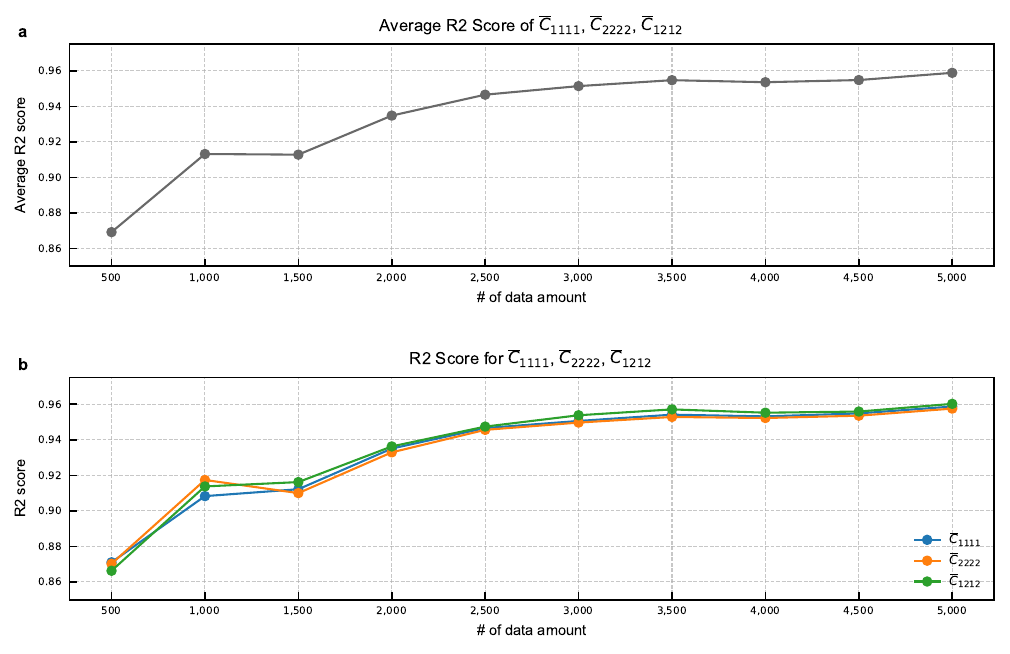}
    \caption{Effect of dataset size on transfer learning performance. (a) Average $\mathrm{R}^2$ scores vs. number of training instances. (b) $\mathrm{R}^2$ scores for individual stiffness components vs the number of training instances.}
    \label{fig:data_size}
\end{figure}

As shown in Figure~\ref{fig:data_size}, performance improves as dataset size increases, but with diminishing returns. Notably, even with just 1,000 samples, the model achieves an $\mathrm{R}^2$ score above 0.91. Beyond 2,500 samples, the performance plateaus, suggesting that relatively small datasets suffice for high-fidelity property prediction.

This data efficiency is particularly important for applications in materials science, where labeled data is often expensive or difficult to obtain. The MMAE’s ability to perform well in low-data regimes makes it a practical and scalable solution for microstructure-informed modeling.

\subsection{Transfer learning for nonlinear behavior prediction}

To assess performance on nonlinear material modeling, we fine-tuned the MMAE to predict IMN parameters $\mathcal{F}(\mathbf{I})$ directly from microstructure images. This approach enables the construction of microstructure-adaptive IMNs, wherein the network parameters are dynamically inferred based on each input microstructure, thereby supporting robust extrapolation to nonlinear mechanical responses under previously unseen loading conditions.

\subsubsection{Offline end-to-end fine-funing}

During offline training, we investigated the effect of varying pre-training masking ratios on transfer learning performance using the short-fiber composite dataset. For each masking ratio, the MMAE was fine-tuned to predict the IMN parameters

\begin{figure}[h]
    \centering
    \includegraphics[width=\linewidth]{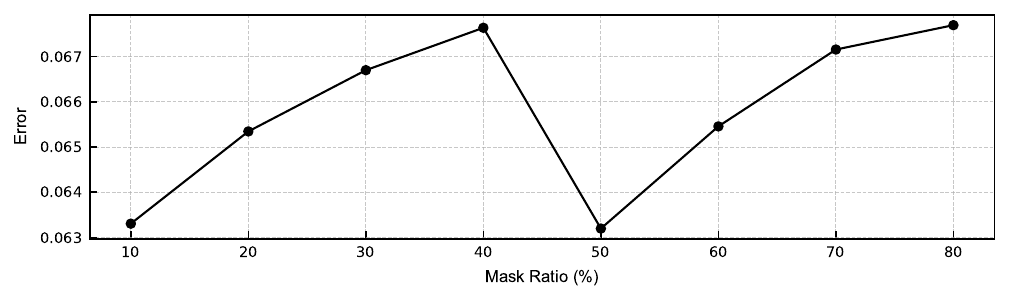}
    \caption{Validation error during offline training for MMAE models pre-trained with different masking ratios.}
    \label{fig:offline_training_performance}
\end{figure}

As shown in Fig.~\ref{fig:offline_training_performance}, validation errors remain consistently low (6.3\%–6.8\%) across all masking ratios, with the best result at 50\%, which we used for online prediction.

This weak dependence on the masking ratio reflects the robustness of end-to-end fine-tuning, where the encoder and prediction head are jointly updated. Unlike linear probing, which relies entirely on frozen representations, end-to-end fine-tuning mitigates the impact of the masking ratio used during pretraining.

\subsubsection{Online prediction}

To assess generalization capability, we evaluated the fine-tuned MMAE-IMN framework on four microstructures that were not part of the 300 RVEs used during offline training. These test cases were subjected to complex loading paths and selected to represent diverse morphological configurations. Despite being trained on a limited dataset, the model accurately predicted the nonlinear mechanical responses of previously unseen microstructures.

\begin{figure}[h]
    \centering
    \includegraphics[width=\linewidth]{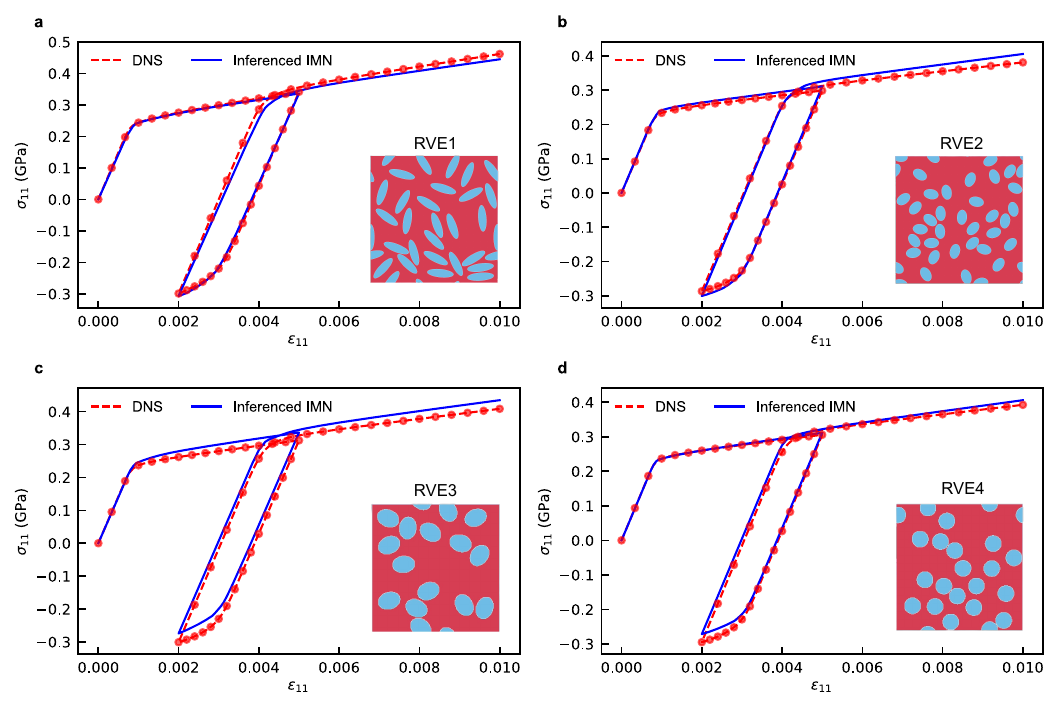}
    \caption{Predicted stress-strain responses for four previously unseen RVEs under complex loading. Insets show the corresponding microstructures, with phase distributions (red: Phase 1, blue: Phase 2).}
    \label{fig:FM-IMN_online_complex_loading}
\end{figure}

Fig.~\ref{fig:FM-IMN_online_complex_loading} presents the predicted stress-strain curves in comparison with DNS results. Each subplot includes an inset depicting the corresponding composite microstructure. The selected cases encompass both short-fiber composites (RVE1--RVE3) and a circular-inclusion configuration (RVE4). The predicted curves closely align with the DNS responses, demonstrating the framework’s robustness across diverse microstructural geometries and nonlinear behaviors.

To quantitatively assess prediction accuracy, we adopt the mean-relative error and max-relative error metrics, following established practice~\cite{huang2022microstructure}. These provide normalized measures of deviation from DNS results:

\begin{equation}
\text{mean-relative error} = 
\frac{ \frac{1}{n} \sum_{i=1}^{n} \left| \sigma_i^{\text{DNS}} - \sigma_i^{\text{pred}} \right| }
     { \max_{i=1,\ldots,n} \left| \sigma_i^{\text{DNS}} \right| }
\label{eq:mean_rel_error}
\end{equation}

\begin{equation}
\text{max-relative error} = 
\frac{ \max_{i=1,\ldots,n} \left| \sigma_i^{\text{DNS}} - \sigma_i^{\text{pred}} \right| }
     { \max_{i=1,\ldots,n} \left| \sigma_i^{\text{DNS}} \right| }
\label{eq:max_rel_error}
\end{equation}

Table~\ref{tab:ss_curve_error_metrics} summarizes the quantitative performance. The mean-relative stress prediction errors are 2.01\%, 2.76\%, 5.40\%, and 2.26\% for RVE1 to RVE4, respectively, while the corresponding maximum-relative errors are 7.34\%, 6.46\%, 7.80\%, and 5.98\%. These results indicate that the MMAE-IMN framework generalizes well to previously unseen microstructures while accurately predicting their nonlinear mechanical responses.

The material properties for Phase 1 and Phase 2 used in the simulations are summarized in Table~\ref{tab:material_properties}. Phase 1 is modeled as an elastoplastic material with isotropic hardening, while Phase 2 is linear elastic.

\begin{table}[htbp]
\centering 
\renewcommand{\arraystretch}{1.3} 
\caption{Material properties for Phase 1 and Phase 2.}
\begin{tabular}{lcccc}
    \hline
    & {Young modulus $E$} & {Poisson ratio $\nu$} & {Yield stress $\sigma_y$} & {Tangent modulus} \\
    \hline
    \textbf{Phase 1} & 200 GPa & 0.3 & 0.2 GPa & 5 GPa \\
    \textbf{Phase 2} & 600 GPa & 0.19 & -- & -- \\
    \hline
\end{tabular}
\label{tab:material_properties}
\end{table}

\begin{table}[htbp]
\centering
\renewcommand{\arraystretch}{1.3}
\caption{Relative stress prediction errors of inferred IMNs.}
\begin{tabular}{lcccc}
    \hline
    Type & RVE1 & RVE2 & RVE3 & RVE4 \\
    \hline
    mean-relative error & 2.01\% & 2.76\% & 5.40\% & 2.26\% \\
    max-relative error & 7.34\% & 6.46\% & 7.80\% & 5.98\% \\
    \hline
\end{tabular}
\label{tab:ss_curve_error_metrics}
\end{table}

\newpage

\subsection{Discussion}

\subsubsection{Training stability of linear probing under random initialization}

To evaluate the robustness of MMAE with respect to random initialization during transfer learning, we performed five independent training runs of the linear probing setup on the short-fiber composite dataset, targeting the prediction of homogenized stiffness components ($\bar{C}_{1111}$, $\bar{C}_{2222}$, $\bar{C}_{1212}$). Each run employed a different random seed, affecting both the model's initial weights and the shuffling of training data.

\begin{figure}[htbp]
\centering
\includegraphics[width=0.95\linewidth]{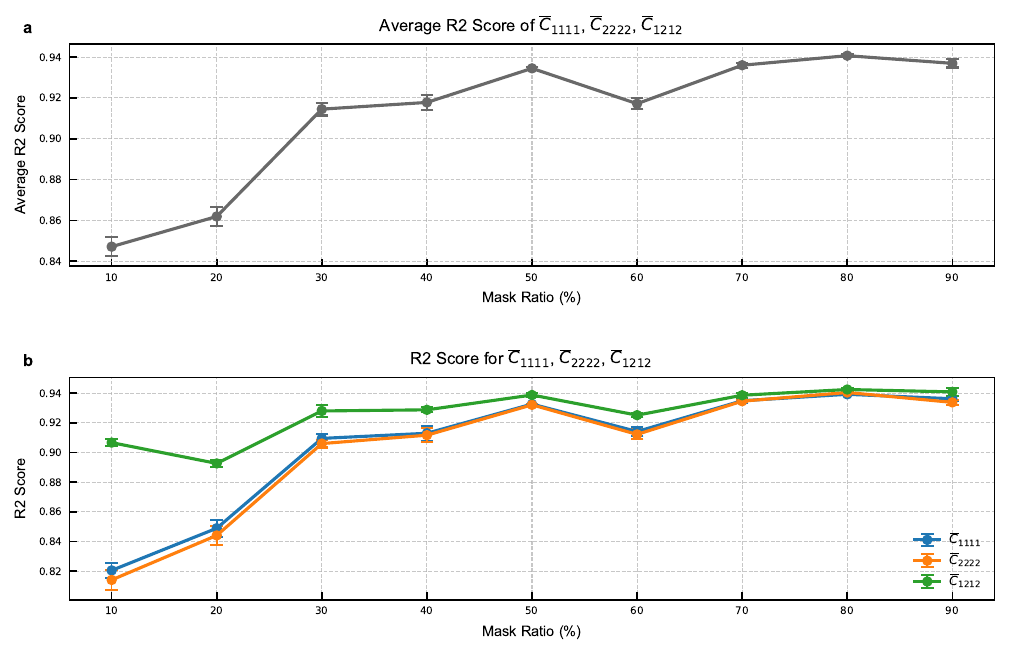}
\caption{Impact of random seed initialization on linear probing results for the short-fiber composite dataset. (a) Mean validation $\mathrm{R}^2$ scores across all stiffness components vs. mask ratio. (b) Component-wise validation $\mathrm{R}^2$ scores. Error bars indicate $\pm1$ standard deviation over five independent runs.}
\label{fig:r3_seed}
\end{figure}

As shown in Fig.~\ref{fig:r3_seed}(a), the average validation $\mathrm{R}^2$ scores across masking ratios exhibit only minor variation, indicating strong consistency across runs. Component-wise performance in Fig.~\ref{fig:r3_seed}(b) further supports this finding, as standard deviations remain low across all evaluated masking ratios. These results confirm that the proposed transfer learning pipeline is largely insensitive to initialization under the linear probing configuration.

\subsubsection{Limitations of 2D pre-training and future directions for 3D extension}

The current MMAE framework is pre-trained exclusively on two-dimensional (2D) synthetic microstructure images. This choice aligns with established practices in composite microstructure analysis, where 2D image-based characterization remains a widely adopted standard due to its computational efficiency, cost-effectiveness, and interpretability under classical stereological assumptions. Prior studies have demonstrated that 2D simulations can reliably estimate effective material properties and capture key statistical features of random heterogeneous media, especially when in-plane behavior dominates or when the microstructure exhibits planar symmetry. For example, Qian et al. demonstrated that 2D finite element models effectively capture size effects in fiber composites under common conditions~\cite{qian2012establishing}. Kaminski and Kazimierczak reported close agreement between 2D and 3D probabilistic homogenization for metallic composites with cylindrical inclusions~\cite{kaminski20142d}. Savvas et al. further validated the use of 2D RVEs by showing accurate elastic property predictions under local volume fraction variation~\cite{savvas2016determination}.

Despite these advantages, 2D representations inherently neglect out-of-plane features. Single cross-sections cannot resolve fiber tilt, curvature, or three-dimensional connectivity, all of which may significantly influence mechanical responses in real materials. As a result, while the model demonstrates strong performance across a range of 2D geometries, its ability to generalize to real 3D morphologies such as clustered, curved, or vertically aligned fibers remains uncertain.

Furthermore, the current training dataset lacks morphological diversity. Within each RVE, inclusions are constrained to have identical shape and size. This simplification reduces structural variability during pre-training, which may further limit the model's capacity to generalize to heterogeneous microstructures.

To address these limitations, future work will extend the MMAE framework to volumetric pre-training using voxelized 3D microstructures that incorporate tilted or serial-section images. These approaches aim to capture spatially coherent features across all three dimensions and expand the training distribution to encompass a broader range of inclusion geometries and spatial arrangements. By lifting the constraints of 2D input, the MMAE framework can be expanded to support more accurate and transferable predictions in structurally complex systems.

\subsubsection{Revisiting masking ratio heuristics for MMAE transfer learning tasks}

In the original MAE literature, it is well established that a high masking ratio during pretraining leads to superior transfer performance, particularly in the linear probing setting. This trend is also observed in our results (Fig.~\ref{fig:linear_probing_SFC} and Fig.~\ref{fig:linear_probing_circle}), where accuracy increases with masking ratio, peaking around 80\%. This is expected, as linear probing freezes the encoder and depends entirely on the quality of representations learned during self-supervised pretraining. Low masking ratios make the reconstruction task too easy, encouraging shortcut learning rather than the abstraction of transferable features.

In contrast, end-to-end fine-tuning continuously updates the encoder during supervised training, allowing the model to refine its latent space based on the downstream objective. This reduces sensitivity to pretraining quality. As shown in Fig.~\ref{fig:fine_tuning_SFC} and Fig.~\ref{fig:fine_tuning_CIC}, the $\mathrm{R}^2$ scores across different masking ratios remain nearly identical, with variations within 0.01. Similarly, Fig.~\ref{fig:offline_training_performance} shows that the corresponding prediction errors differ by less than 0.5\%. These results confirm the robustness of end-to-end adaptation, but deviate from trends commonly observed in vision benchmarks, where high masking ratios yield measurable improvements even under fine-tuning~\cite{he2022masked}.

We hypothesize that this discrepancy arises from two factors:  
\begin{enumerate}
    \item The downstream task—predicting homogenized stiffness matrices—is relatively simple compared to complex visual recognition tasks, so the model can compensate for suboptimal pretraining through supervised fine-tuning.

    \item The overall dataset complexity may not be high enough to magnify differences in representation quality, even though linear probing already revealed some sensitivity to masking ratio.
\end{enumerate}

These findings suggest that pretraining heuristics established in computer vision, such as the use of high masking ratios, may not directly generalize to scientific domains. Domain-specific foundation models may require distinct pretraining strategies, and we highlight this issue as an open question for future research.

\newpage

\section{Conclusions}\label{sec4}
This work proposed the MMAE, a foundation model specifically designed for composite materials, aimed at addressing key challenges in data scarcity and transferability within materials science. By leveraging self-supervised learning on a large dataset of short-fiber composite microstructures, the MMAE learned robust and generalizable latent representations, forming a strong basis for a wide range of downstream tasks.

In the first transfer learning scenario, the MMAE was fine-tuned to predict homogenized stiffness components, achieving an $\mathrm{R}^2$ score of up to 0.959, even with limited labeled data. This demonstrates the model's ability to perform accurate property prediction in data-constrained regimes. In the second scenario, the MMAE was fine-tuned to infer IMN parameters, enabling accurate extrapolation of nonlinear mechanical behavior across unseen microstructures and loading conditions.

These results underscore the promise of foundation models in computational materials science, providing scalable, data-efficient tools for microstructure-informed property prediction. Future work will extend the MMAE framework to more complex systems such as three-dimensional composites and polycrystalline aggregates, as well as integrate experimental datasets to enhance its applicability in real-world materials design and characterization.

\appendix

\section{Microstructure generation}
\label{sec:Microstructure_generation:appendix}

\begin{algorithm}
\caption{Synthetic RVE Generation}
\label{alg:rve_generation}
\begin{algorithmic}[1]
\Function{GenerateRVEs}{$N_{\text{samples}}$}
    \State $\mathcal{S} \gets$ \Call{LatinHypercubeSample}{$N_{\text{samples}},\ (N_p{:}\ 15\text{--}35,\ A_r{:}\ 1\text{--}4,\ v_f{:}\ 10\%\text{--}40\%)$}

    \For{each $(N_p, A_r, v_f)$ in $\mathcal{S}$}
        \State $(a, b) \gets$ \Call{ComputeEllipseAxes}{$N_p, v_f, A_r$}
        \State RVE $\gets$ \Call{InitializeDomain}{size}
        \For{$j = 1$ to $N_p$}
            \While{true}
                \State $(x, y, \theta) \gets$ \Call{SamplePositionAndOrientation}{}
                \State $p \gets$ \Call{CreateEllipse}{$a, b, \theta$}
                \If{\Call{OverlapsWithOthers}{$p$, RVE}}
                    \State \Call{Remove}{$p$}
                    \State \textbf{continue}
                \EndIf
                \If{\Call{CrossesBoundary}{$p$}}
                    \State \Call{ApplyPeriodicImage}{$p$}
                    \If{\Call{OverlapsWithOthers}{$p$, RVE}}
                        \State \Call{Remove}{$p$}
                        \State \textbf{continue}
                    \EndIf
                \EndIf
                \State \textbf{break}
            \EndWhile
        \EndFor
        \State \Call{SaveRVE}{RVE}
    \EndFor
\EndFunction
\end{algorithmic}
\end{algorithm}

\newpage

\section*{Acknowledgements}
This work is supported by the National Science
and Technology Council, Taiwan, under Grant 111-2221-E-002-054-
MY3 and 112-2221-E-007-028. We are grateful for the computational
resources and support from the NTUCE-NCREE Joint Artificial Intelligence Research Center and the National Center of High-performance Computing (NCHC). 

\section*{Declarations}
The authors declare that there are no competing interests.

\section*{Data Availability}
The datasets and trained models used in this study are publicly available at the Zenodo repository: \url{https://doi.org/10.5281/zenodo.14062123}.

\section*{Code Availability}
The code used in this study can be accessed at the following GitHub repository: \url{https://github.com/BerryWei/Material_mask_autoencoder}.

\section*{Author Contributions}
Conceptualization and Project Administration: All authors. Investigation and methodology: Ting Ju Wei. Writing—Original Draft: Ting Ju Wei. Review \& Editing: All authors. Resources and Funding Acquisition: Chuin-Shan Chen. 
These author contributions are defined according to the CRediT contributor roles taxonomy.

\bibliographystyle{unsrt}  
\bibliography{references}






\end{document}